\documentclass[twocolumn,showpacs,showkeys,preprintnumbers,amsmath,amssymb]{revtex4-2}
\usepackage[latin9]{inputenc}
\usepackage{amsmath}
\usepackage{amssymb}
\usepackage{graphicx}
\usepackage{ulem}
\usepackage{xcolor}
\usepackage{dcolumn}
\usepackage{soul}
\setstcolor{blue}

\begin{document}
\title{Novel predator-prey model admitting exact analytical solution}
\date{\today}
\author{G. Kaniadakis}
\email{giorgio.kaniadakis@polito.it}
\affiliation{Department of Applied Science and Technology, Politecnico di Torino, \\
Corso Duca degli Abruzzi 24, 10129 Torino, Italy}
\begin {abstract}
The Lotka-Volterra predator-prey model still represents the paradigm for the description of the competition in population dynamics. Despite its extreme simplicity, it does not admit an analytical solution, and for this reason, numerical integration methods are usually adopted to apply it to various fields of science. The aim of the present work is to investigate the existence of new predator-prey models sharing the broad features of the standard Lotka-Volterra model and, at the same time, offer the advantage of possessing exact analytical solutions. To this purpose, a general Hamiltonian formalism, which is suitable for treating a large class of predator-prey models in population dynamics within the same framework, has been developed as a first step. The only existing model having the property of admitting a simple exact analytical solution, is identified within the above class of models. The solution of this special predator-prey model is obtained explicitly, in terms of known elementary functions, and its main properties are studied. Finally, the generalization of this model, based on the concept of power-law competition, as well as its extension to the case of $N$-component competition systems, are considered.
\end {abstract}

\maketitle

\section{Introduction}

The standard Lotka-Volterra (LV) predator-prey model is described by the bilinear first-order coupled differential equations
\begin{eqnarray}
&&\frac{d\,x_1}{d\,t}= c_{12} \,x_1\,x_2 -c_{11} \,x_1 \ \ , \label{LVI1} \\
&&\frac{d\,x_2}{d\,t}= -c_{21} \, x_1 \, x_2 +c_{22} \, x_2 \ \ ,
\label{LVI2}
\end{eqnarray}
where the positive functions $x_1=x_1(t)$ and $x_2=x_2(t)$ represent the predator and the pray populations, respectively, while $c_{ij}$ are positive constants describing the interaction between the two populations.

The above deterministic model was created in the third decade of the twentieth century in the field of ecology \cite{Lotka,Volterra}. Almost a century after it was first proposed it still attracts the interest of the scientific community, and intense research activity has been carried out on this model up to the present day \cite{Llibrea,Ghasemabadi,Jiang,Morrison,Palombi,Martins,Bunin,Chen,Lin,Skvortsov}.

Ecology undoubtedly represents the field in which the LV model has been used the most and in a systematic way, but its versatility has allowed it to be used for the study of an ever increasing variety of physical, natural and artificial systems, such as in plasma physics \cite{Plasma}, in spin-wave patterns \cite{Spin}, in the formation of crystallization fronts \cite{Crystal}, in multimode dynamics in optical systems \cite{Optics}, in neural networks \cite{Networks} etc. A very large, but not exhaustive, list of applications of the LV model can be found in Refs. \cite{Book1,Book2}.

The undisputed success of the LV model has not prevented researchers from going further and proposing generalizations of the model in order to describe some empirically observed phenomenologies, especially in the field of ecology and biology \cite{Rosenzweig,Albrecht,Freedman,Arditi,Boudjellaba}. A huge debate is currently ongoing in ecology on how to model predation, and the discussion on how the predator's consumption rate (known as functional response) can be influenced by the predator and the prey population densities, has been stimulated by some works \cite{SongSaavedra,AbramsGinzburg,HattonMacCann,O'Dwyer,Berryman}. In prey-dependent models the functional response is independent on predator population size while in the  predator-dependent models the functional response depends of both predator and prey population sizes. Special case of predator-dependent models are the so called ratio-depended models (ratio of prey population size to predator population size) promoted by Arditi and Ginzburg. A critical discussion on the principal predator-prey models and limits of their validity can be found in ref. \cite{AbramsGinzburg}. Mathematical generalizations of the LV model has recently been considered in the framework of fractional calculus \cite{Ahmed,Das}.

An old but very efficient and simple tool that is used in the construction of ordinary differential equations, and which is suitable for describing the evolution of populations, is undoubtedly the formalism of power-law functions. It is worth noting that the first equation, describing the evolution of a population, proposed by Verhulst in 1838, contains a term governing the saturation of the population, which is a second-degree power-law of the population function. In the last two decades of the twentieth century, the formalism of power laws was systematically employed to propose population evolution models described by coupled, first-order, ordinary nonlinear differential equation systems \cite{Hernandez}. This formulism was first introduced in theoretical biochemistry \cite{SaVo,Savageau,Voit,VoitSavageau}. In ecology, an important generalization of the LV system, involving the power-law formalism, was proposed in 1988 by Brenig \cite{Brenig}.

Despite its apparent simplicity, the LV model is far from being understood. It does not admit analytical solution and, for this reason, its solution is obtained  numerically or in an approximate way by considering its linearized version. These difficulties have not prevented interest in the LV model, which still represents the paradigm for describing the phenomenon of predator-prey competition. However,  a common feature of all the models that generalize the LV one is that they do not lead to analytical solutions as well.

Knowledge of the possible analytical solution, in terms of known elementary functions of an evolution model of competing populations is important from, and not only the practical point of view because of the possibility of conducting quick comparison with empirical data. The existence of a simple analytical solution is also extremely important for a model, from a theoretical point of view, as it allows us to better understand the nature of the competition mechanism. This ultimately gives an added value to the model itself, as it makes it more transparent.

The aim of the present work is to propose a novel predator-prey model that captures the main features and symmetries of the standard LV model and, at the same time, has the additional property of admitting a simple analytical solution, in terms of known elementary functions.
Furthermore, the generalization of this model, in the context of the power-law function formalism, yields a second, more general, model that presents a simple Hamiltonian form, suitable for numerical evaluation.

The paper is organized as follows: First, we introduce a general class of Hamiltonian predator-prey models, where the competition terms are expressed through two arbitrary functions and its main properties are studied. In particular, we obtain the first integral of motion of this class of models and focus on their Hamiltonian formalism.  Subsequently, we identify the only existing model, within this general class of models, admitting an exact analytical solution and obtain it in explicit form. Then, as a first extension of this integrable model, a class of models involving power-law competition rates, is proposed. Finally a further extension of the formalism developed for the here proposed predator-prey system, in the case of an arbitrary number of interacting populations, is also studied.

\section{A general class of  Predator-Prey  Models}

Four fundamental properties of the standard LV model immediately emerge from a direct inspection of the two equations expressing the rate of change of the two predator-prey populations:

i) The rate of change of each population is assumed to be the difference between rate of growth and rate of loss.

ii) The competition term plays the growth rate role for the predator and of loss rate role for the prey, and it is obtained by considering the product of the contribution of the two populations.

iii) The contribution of the two populations in the expressions of the rate terms is quantified through two characteristic functions.

iv) The two characteristic functions of the competing populations are assumed to be linear functions of the related populations  $ f_i (x_i) = x_i $.

Hereafter, we consider a class of predator-prey models for which only the first three properties of the standard LV model continue to apply i.e. i), ii) and iii). The fourth property is released, so that the new class of predator-prey models is obtained, starting from the equations that define the LV model and after substituting $x_i \rightarrow f_i(x_i)$, in all the rate term, in the right hand side of Eqs. (\ref{LVI1}) and (\ref{LVI2}), thus obtaining
\begin{eqnarray}
&&\frac{d\,x_1}{d\,t}= c_{12} \, f_1(x_1)\,f_2(x_2) -c_{11} \, f_1(x_1) \ \ , \label{LVII1}  \\
&&\frac{d\,x_2}{d\,t}= -c_{21} \, f_1(x_1) \, f_2(x_2) +c_{22} \, f_2(x_2) \ \ .
\label{LVII2}
\end{eqnarray}
The above coupled differential equations describe a class of predator-prey models that capture many of the features of the standard LV one. This class of models is very general, because of the arbitrariness of the two characteristic functions $ f_i (x_i) $ that obey the conditions $ f_i (0)=0 $ and $ d\,f_i (x_i)/d\,x_i>0 $.

The two coupled first order differential Eqs. (\ref{LVII1}) and (\ref{LVII2}) can be easily uncoupled obtaining the two second order differential equations for the predator and prey populations. First we introduce the transformed populations $w_i=w_i(t)$ defined through  $w_i=f_i(x_i)$ with $i=1,2$ and the auxiliary functions $\Lambda_i(w)$ according to
\begin{eqnarray}
\frac{d \,\Lambda_i(w)}{d\, w} = \frac{1}{w\,f'_i(f_i^{-1}(w))} \ \ ,
\label{LVA}
\end{eqnarray}
$f'_i(x_i)$ and $f_i^{-1}(x_i)$ indicating the derivative and the inverse function of $f_i(x_i)$ respectively. Then the introduction of the two non-linear first order differential operators
\begin{eqnarray}
{\cal D}_{ij}(w)=\frac{c_{ii}}{c_{ij}} +\frac{1}{c_{ij}}\,\frac{d \,\Lambda_i(w)}{d\, w} \, \frac{d\, w}{d\,t} \ \ ,
\label{LVB}
\end{eqnarray}
permits us to write the predator evolution equation in the form of the following non-linear second-order differential equation
\begin{eqnarray}
 {\cal D}_{21} \! \left ({\cal D}_{12}(w_1)\right) +w_1= \lambda_{21} \ \ ,
\label{LVB}
\end{eqnarray}
with $\lambda_{ij}= 2\, c_{ii}/c_{ij}$. The prey evolution equation follows from the predator one by exchanging the indexes $1 \leftrightarrow 2 $ and inverting time $t \rightarrow - t$. It is remarkable that the dynamics of the predator-pray system is univocally fixed by the forms of the two functions $f_i(x)$ or equivalently by the forms of the two functions $\Lambda_i(w)$.

\section{Hamiltonian formalism}

To introduce the Hamiltonian formalism of the above  general class of models we first obtain the first integral of motion  of the system. By direct comparison of Eqs. (\ref{LVII1}) and (\ref{LVII2}),  the differential equation follows as
\begin{eqnarray}
\frac{d\,x_2}{d\,x_1}= -\frac{f_2(x_2)}{f_1(x_1)}\, \frac{c_{21} \, f_1(x_1) -c_{22}}{c_{12} \, \,f_2(x_2) -c_{11}}   \ \ .
\label{LVII3}
\end{eqnarray}
After separation of the variables
\begin{equation}
\left (c_{12} -\frac{c_{11}}{f_2(x_2)} \right) dx_2 + \left (c_{21} -\frac{c_{22}}{f_1(x_1)} \right) dx_1 =0 \ ,
\label{LVII4}
\end{equation}
and integration, the first integral of motion is obtained in the form
\begin{eqnarray}
H= c_{21} \, x_1 - c_{22}\,F_1(x_1)  + c_{12}\, x_2 - c_{11}\,F_2(x_2)  \ , \ \
\label{LVII5}
\end{eqnarray}
where $H$ is the integration constant while the $F_i(x_i)$ functions are defined according to
\begin{eqnarray}
\frac{d F_i(x_i)}{dx_i}=\frac{1}{f_i(x_i)} \ \ .
\label{LVII6}
\end{eqnarray}

The quantity $H$, can be viewed as a function of the two variables $x_i$ i.e. $H=H(x_1, x_2)$ and, after taking into account the expression of $H$ given by Eq. (\ref{LVII5}), the definition (\ref{LVII6}), the evolution equations (\ref{LVII1}) and (\ref{LVII2}) it follows that
\begin{equation}
\frac{d\, H}{d\,t} =\frac{\partial \, H}{\partial\,x_1}\,\frac{d\, x_1}{d\,t} +\frac{\partial\, H}{\partial\,x_2}\frac{d\, x_2}{d\,t} = 0 \ ,
\label{LVII7}
\end{equation}
so that we can conclude that $H$ is a conserved quantity whose value $H_0$, depends on the initial conditions. Thus, equation $H(x_1, x_2)=H_0$ defines, in the phase space i.e. the $x_1 x_2$ plane, a level curve of the function $H(x_1, x_2)$ that is the orbit of the system.

By employing the above introduced functions $F_i(x_i)$, the predator-prey evolution Eqs. (\ref{LVII1}), (\ref{LVII2}) can be written in the form
\begin{eqnarray}
&&\frac{d\,F_1(x_1)}{d\,t}=  c_{12} \,f_2(x_2) -c_{11}\ \ , \label{LVII9} \\
&&\frac{d\,F_2(x_2)}{d\,t}= c_{22}- c_{21} \, f_1(x_1)  \ \ .
\label{LVII10}
\end{eqnarray}
The introduction of the transformation
\begin{eqnarray}
y_i=F_i(x_i) \ ,
\label{LVII11}
\end{eqnarray}
appears natural at this point. The evolution equations for the transformed populations $y_i=y_i(t)$ become
\begin{eqnarray}
&&\frac{d\,y_1}{d\,t}=  c_{12} \,\phi_2(y_2)-c_{11} \  , \label{LVII12} \\
&&\frac{d\,y_2}{d\,t}= - c_{21} \, \phi_1(y_1) + c_{22}   \ ,
\label{LVII13}
\end{eqnarray}
with $\phi_i(y_i)=f_i(\Phi_i(y_i))$ and $\Phi_i(y_i)= F_i^{-1}(y_i)$ being the inverse function of $F_i(x_i)$.

As a consequence, the inverse transformations of the ones defined through Eq. (\ref{LVII11}) assume the form
\begin{eqnarray}
x_i=\Phi_i(y_i) \ ,
\label{LVII16}
\end{eqnarray}
so that the conserved quantity $H$ of the system governed by evolution Eqs. (\ref{LVII1}) and (\ref{LVII2}), if expressed in terms of the transformed populations $y_i$, takes the form
\begin{equation}
H= c_{21}\,\Phi_1(y_1) -c_{22} \, y_1 + c_{12}\,\Phi_2(y_2) - c_{11}\, y_2   \ .
\label{LVII17}
\end{equation}

An alternative expression of the function $\phi_i(y_i)$ follows easily, after taking into account Eqs. (\ref{LVII6}) and (\ref{LVII11}). It thus obtains
\begin{eqnarray}
\phi_i(y_i)\!\!\!\!&&=f_i(\Phi_i(y_i))=f_i(x_i)=\frac{1}{\frac{d\, F_i(x_i)}{d\, x_i}}=\frac{d\, x_i}{d\, y_i}  \ ,  \ \ \ \ \ \ \ \ \ \label{LVII18}
\end{eqnarray}
and after taking into account Eq. (\ref{LVII16}), it follows
\begin{eqnarray}
\phi_i(y_i)=\frac{d\, \Phi_i(y_i)}{d\, y_i} \ .
\label{LVII20}
\end{eqnarray}

The thus obtained direct differential link between the $\phi_i(y_i)$ and $\Phi_i(y_i)$ functions, permits us to write evolution equations (\ref{LVII12})) and (\ref{LVII13}), for the transformed populations $y_i$ in canonical form
\begin{eqnarray}
&&\frac{d\,y_1}{d\,t}=\frac{\partial \, H}{\partial \, y_2}  \ ,  \label{LV21}\\
&&\frac{d\,y_2}{d\,t}=-\frac{\partial \, H}{\partial \, y_1}  \ .
\label{LV22}
\end{eqnarray}
We can then conclude that the conserved quantity $H$, is the Hamiltonian of the transformed canonical system $\{y_i(t)\}$. In other words, transformation (\ref{LVII11}) maps the real system $\{x_i(t)\}$ into the canonical one $\{y_i(t)\}$.

It is remarkable that, according to the present formalism, any predator-prey system described by evolution equations (\ref{LVII1}) and (\ref{LVII2}), is to associated one and only one canonical system, whose dynamics is governed by evolution equations (\ref{LVII12}) and (\ref{LVII13}). The nature of the predator-prey competition is fixed by the two functions $f_i(x_i)$. Furthermore, the functions $f_i(x_i)$ univocally determine the link between the real and the canonical system by fixing the functions $F_i(x_i)$ and $\Phi_i(x_i)$ and also the functions $\phi_i(x_i)$ which govern the competition dynamics for the canonical system, according to the scheme
\begin{eqnarray}
f_i(x_i)\Longleftrightarrow F_i(x_i) \Longleftrightarrow \Phi_i(y_i)\Longleftrightarrow \phi_i(y_i) \ .
\label{LVII23}
\end{eqnarray}

It is worth noting that the transformations that allows the real system to be mapped to the canonical system, i.e. the functions $ \Phi_i (y_i) $ can be obtained directly from the functions $ f_i (x_i) $. Indeed, Eq. (\ref{LVII20}), after taking into account that $\phi_i(y_i)=f_i(\Phi_i(y_i))$, assumes the form
\begin{eqnarray}
\frac{d\,\Phi_i(y_i)}{d\,y_i}=f_i(\Phi_i(y_i)) \ .
\label{LVII24}
\end{eqnarray}
Finally, from the last differential equation and after taking into account Eq. (\ref{LVII20}), it easily follows also the direct link between the functions $f_i(x_i)$ and $\phi_i(y_i)$
\begin{eqnarray}
\phi_i(y_i)=f_i\left( \,\int_0^{y_i}\!\phi_i(w) \, dw \right) \ .
\label{LVII25}
\end{eqnarray}

Starting from a given predator-prey system, in which $f_i(x_i)$ is fixed, the two latter equations can be employed to obtain the associated Hamiltonian system. For instance, in the case of the standard LV system, described by Eqs. (\ref{LVI1}) and (\ref{LVI2}) where $f_i(x_i)=x_i$,  we obtain $\Phi_i(z)=\exp(z)$ and $\phi_i(z)=\exp(z)$, by employing Eqs. (\ref{LVII24}) and (\ref{LVII25}) respectively. The evolution equations for the canonical counterpart of the standard LV model, then reads
\begin{eqnarray}
&&\frac{d\,y_1}{d\,t}=  c_{12} \,\exp(y_2)-c_{11} \  , \label{LVII26} \\
&&\frac{d\,y_2}{d\,t}=c_{22} - c_{21} \, \exp(y_1)    \ .
\label{LVII27}
\end{eqnarray}

The introduction of the Hamiltonian formalism  is not of great utility for the standard LV model, from the computational point of view. The system, also in its canonical form, already known in literature \cite{FVerhulst}, continues to not admit any analytical solution. Interestingly, the Hamiltonian formalism turns out to be very useful for the construction of a new, exactly solvable, predator-prey model, as will be seen below.

\section{The new exactly solvable model}

Hereafter, we are interested in identifying the predator-prey model among the infinity of models described by Eqs. (\ref{LVII1}) and (\ref{LVII2}), if it exists,  that admits an explicit analytical solution. This requirement can be imposed on the Hamiltonian system so that, starting from a solvable Hamiltonian system, we can go back to identify the corresponding real system by exploiting the transformation that links the two systems. The existence of an analytical solution for the Hamiltonian system described by Eqs. (\ref{LVII12}) and (\ref{LVII13}), can be imposed by requiring it to be linear, i.e.,
\begin{eqnarray}
\phi_i(y_i)=2\, y_i \ .
\label{LVIII1}
\end{eqnarray}
The non-essential multiplicative constant has been chosen equal to $ 2 $, in such a way that the transformation allowing to pass from the Hamiltonian system to the real one has the following simple form:
\begin{eqnarray}
\Phi_i(y_i)= y_i^2 \ ,
\label{LVIII2}
\end{eqnarray}
as imposed by Eq. (\ref{LVII20}).

Starting from the above expression of the functions $\phi_i(y_i)$ or $\Phi_i(y_i)$ and by employing Eqs. (\ref{LVII25}) and (\ref{LVII24}), respectively, the functions $f_(x_i)$ follows immediately. For instance, after substituting the expression of $\Phi_(y_i)$, as given by Eq. (\ref{LVIII2}), into Eq. (\ref{LVII24}), it obtains the functional equation $2\,y_i=f_i(y_i^2)$, which yields
\begin{eqnarray}
f_i(x_i)= 2\,\sqrt{x_i} \ .
\label{LVIII3}
\end{eqnarray}
From Eq. (\ref{LVII6}) and the expression of $f_i(x_i)$, it follows that
\begin{eqnarray}
F_i(x_i)= \sqrt{x_i} \ .
\label{LVIII4}
\end{eqnarray}

At this point, after posing  $\gamma_{ii}=2 \, c_{ii}$, $\gamma_{ij}=4 \, c_{ij}$ with $i=1,2$, $j=1,2$ and $i\neq j$, for simplicity,  we can write the evolution equations for the real system
\begin{eqnarray}
&&\frac{d\,x_1}{d\,t}= \gamma_{12} \, \sqrt{x_1}\,\sqrt{x_2} -\gamma_{11} \, \sqrt{x_1} \ \ , \label{LVIII7} \\
&&\frac{d\,x_2}{d\,t}= -\gamma_{21} \, \sqrt{x_1} \, \sqrt{x_2} +\gamma_{22} \, \sqrt{x_2} \ \ ,
\label{LVIII8}
\end{eqnarray}
and for the corresponding Hamiltonian system
\begin{eqnarray}
&&\frac{d\,y_1}{d\,t}=  \frac{\gamma_{12}}{2}\, y_2 - \frac{\gamma_{11}}{2} \ \ , \label{LVIII9} \\
&&\frac{d\,y_2}{d\,t}= -\frac{\gamma_{21}}{2}\, y_1 + \frac{\gamma_{22}}{2}  \ \ .
\label{LVIII10}
\end{eqnarray}
together with the quadratic transformation
\begin{eqnarray}
x_i=y_i^2 \ ,
\label{LVIII11}
\end{eqnarray}
that links the two systems.

It is remarkable that evolution equations (\ref{LVIII7}) and (\ref{LVIII8}) describe a new simple but different from the standard LV predator-prey model. Interestingly, this new model captures the main features of the LV model and additionally has the property to admit an exact analytical solution, which can be expressed in terms of known elementary functions. It is important to note that the discovery of the above integrable model does not exclude the existence of other integrable models within the class described by Eqs (\ref{LVII1}) and (\ref{LVII2}).

The conserved quantity $H$ assumes, for the two systems, the forms
\begin{eqnarray}
H= \frac{\gamma_{21}}{4} \, x_1 - \frac{\gamma_{22}}{2}\, \sqrt{x_1}  + \frac{\gamma_{12}}{4}\, x_2 - \frac{\gamma_{11}}{2}\,\sqrt{x_2}  \ , \ \
\label{LVIII12}
\end{eqnarray}
and
\begin{equation}
H=\frac{\gamma_{21}}{4}\,y_1^2-\frac{\gamma_{22}}{2} \, y_1 + \frac{\gamma_{12}}{4}\,y_2^2   - \frac{\gamma_{11}}{2}\, y_2  \ .
\label{LVIII13}
\end{equation}

The above expressions of $H$ defines the orbits of the system in the two $x_1x_2$ and $y_1y_2$ phase spaces. After some tedious but simple algebra, and after posing
\begin{equation}
y_{1c}=\frac{\gamma_{22}}{\gamma_{21}} \ \ \ , \ \ \ y_{2c}=\frac{\gamma_{11}}{\gamma_{12}}  \ ,
\label{LVIII14}
\end{equation}
\begin{equation}
h= H + \frac{\gamma_{11}^2}{4\,\gamma_{12}} + \frac{\gamma_{22}^2}{4\,\gamma_{21}}  \ ,
\label{LVIII15}
\end{equation}
and
\begin{equation}
a_1=\sqrt{ \frac{4\,h}{\gamma_{21}}} \ \ \ , \ \ \ a_2=\sqrt{ \frac{4\,h}{\gamma_{12}}  }   \ , \label{LVIII16}
\end{equation}
Eq. (\ref{LVIII13}), which defines the orbit of the Hamiltonian system in the $y_1y_2$ plane, assumes the form
\begin{equation}
\frac{(y_1-y_{1c})^2}{a_1^2} + \frac{(y_2-y_{2c})^2}{a_2^2}=1    \ ,
\label{LVIII17}
\end{equation}
thus indicating that the orbit is an ellipse.

The solution of the linear system of Eqs. (\ref{LVIII9}) and (\ref{LVIII10}) is given by
\begin{eqnarray}
y_1= y_{1c} + a_1 \, \sin(\omega \, t + \theta_0) \ , \label{LVIII18}\\
y_2= y_{2c} + a_2 \, \cos(\omega \, t +\theta_0)  \ , \label{LVIII19}
\end{eqnarray}
where the frequency $\omega$ is given by
\begin{eqnarray}
\omega=\frac{1}{2}\, \sqrt{\gamma_{12} \gamma_{21}} \ \ , \label{LVIII20}
\end{eqnarray}
while $\theta_0$ is related to the initial conditions, according to
$\sin (\theta_0)=(y_{10}-y_{1c})/a_1$
and $\cos (\theta_0)=(y_{20}-y_{2c})/a_2$.

After substitution of the expressions of $\theta_0$, $y_{1c}$, $y_{2c}$, $a_1$ and $a_2$ in Eqs. (\ref{LVIII18}) and (\ref{LVIII19}), the time evolution of the Hamiltonian populations $y_i$ can be obtained, in terms of the initial conditions $y_{i0}$ and the model parameters $\gamma_{ij}$.
Finally, from the quadratic transformation given by Eq. (\ref{LVIII11}), we immediately obtain the time evolution of $x_i$
\begin{eqnarray}
 x_1= \bigg [\!\!\!\!\!&& \frac{\gamma_{22}}{\gamma_{21}}+ \left( \sqrt{x_{10}}-\frac{\gamma_{22}}{\gamma_{21}}\right)\, \cos (\omega \, t) \nonumber \\ &&+\, \sqrt{\frac{\gamma_{12}}{\gamma_{21}}} \, \left(\sqrt{x_{20}}-\frac{\gamma_{11}}{\gamma_{12}}\right)\,\sin (\omega \, t) \bigg ] ^2 , \label{LVIII25} \\
 x_2= \bigg [\!\!\!\!\!&& \frac{\gamma_{11}}{\gamma_{12}}+ \left(\sqrt{x_{20}}-\frac{\gamma_{11}}{\gamma_{12}}\right)\, \cos (\omega \, t) \nonumber \\ &&-\, \sqrt{\frac{\gamma_{21}}{\gamma_{12}}} \, \left(\sqrt{x_{10}}-\frac{\gamma_{22}}{\gamma_{21}}\right)\,\sin (\omega \, t)\bigg ] ^2 , \label{LVIII26}
\end{eqnarray}
The above obtained functions, $x_i=x_i(t)$, represent the general exact analytical solution of the predator-prey system described by evolution equations (\ref{LVIII7}) and (\ref{LVIII8}).
It is remarkable that the competition rates in the present model are described through the characteristic sub-linear functions $f_i(x_i)=\sqrt{x_i}$ which substitute the linear functions $f_i(x_i)=x_i$ appearing in the standard LV model. This simple substitution is sufficient to generate a predator-prey model admitting an exact analytical solution.

The existence of the above constructed diffeomorphism linking the real non-linear dynamical system (exhibiting an eliptic fixed point)  and the Hamiltonian linear system (with a center), is guaranteed by the  Poincare-Bendixson theorem. This diffeomorphism is obtained here together with the solution of the real dynamic system and its frequency in closed form.
For the standard LV system, the frequency and period can only be estimated numerically after evaluating a complicated integral \cite{Waldvogel}. The need to have the expression of the period of the standard system of LV in a closed form is strongly felt and has led towards the proposal of asymptotic analytic expressions that provide only approximate estimations \cite{Oshime}.
This highlights the usefulness of having in closed form the solution and the frequency of the system described by the model proposed here, especially in view of its applications.

\section{Power-law predator-prey model}

An important predator-prey model is obtained when $f_i(x_i)$ are  power-law functions i.e. $f_i(x_i)\propto x_i^{\alpha_i}$, with $\alpha_i >0$. In such a case, $F_i(x_i)$, $\Phi_i(x_i)$ and $\phi_i(x_i)$ also become power-law functions i.e.
\begin{eqnarray}
&&f_i(x_i)=\frac{1}{1-\alpha_i}\,x_i^{\alpha_i} \ \ , \label{LVIV1} \\
&&F_i(x_i)=x_i^{1-\alpha_i}   \ \ , \label{LVIV2} \\
&&\Phi_i(y_i)= y_i^{\frac{1}{1-\alpha_i}} \ \ , \label{LVIV3} \\
&&\phi_i(y_i)=\frac{1}{1-\alpha_i}\, y_i^{\frac{\alpha_i}{1-\alpha_i}} \ \ . \label{LVIV4}
\end{eqnarray}

After posing $\gamma_{ii}=c_{ii}/(1-\alpha_i)$, $\gamma_{ij}=c_{ij}/(1-\alpha_i)(1-\alpha_j)$, with $j\neq i$ and $i=1,2$, $j=1,2$, Eqs, (\ref{LVIII7}) and (\ref{LVIII8}), that define the competition model, assume a particularly simple form
\begin{eqnarray}
&&\frac{d\,x_1}{d\,t}= \gamma_{12} \, x_1^{\alpha_1} \,x_2^{\alpha_2} -  \gamma_{11}\,x_1^{\alpha_1} \ , \label{LVIV9} \ \ \ \ \ \ \ \\
&&\frac{d\,x_2}{d\,t}= -\gamma_{21} \, x_1^{\alpha_1} \,x_2^{\alpha_2}  + \gamma_{22}\,x_2^{\alpha_2} \ , \label{LVIV10} \ \ \ \ \ \ \ \
\end{eqnarray}
while the first integral of motion becomes
\begin{eqnarray}
H\!\!\!\!\!\!&&= (1-\alpha_1)(1-\alpha_2)\gamma_{21} \, x_1 - (1-\alpha_2)\gamma_{22}\,x_1^{1-\alpha_1}  \nonumber \\ &&+ \, (1-\alpha_1)(1-\alpha_2)\gamma_{12}\, x_2 - (1-\alpha_1)\gamma_{11}\,x_2^{1-\alpha_2} \ . \nonumber \\
\label{LVIV11}
\end{eqnarray}

The associated canonical system can be introduced by considering the transformation $y_i=x_i^{1-\alpha_i}$
so that, after posing $\beta_i=\alpha_i/(1-\alpha_i)$ and $\eta_{ij}=\gamma_{ij}/(1+\beta_i)$,
the evolution equations become
\begin{eqnarray}
&&\frac{d\,y_1}{d\,t}=  \eta_{12} \, y_2^{\beta_2} - \eta_{11} \  , \label{LVIV15} \\
&&\frac{d\,y_2}{d\,t}= -\eta_{21} \, y_1^{\beta_1} + \eta_{22}    \ ,
\label{LVIV16}
\end{eqnarray}
while the Hamiltonian of the system assumes the form
\begin{equation}
H= \frac{\eta_{21}}{1+\beta_1} \, y_1^{1+\beta_1} - \eta_{22}\,y_1  + \frac{\eta_{12}}{1+\beta_2} \, y_2^{1+\beta_2} - \eta_{11}\,y_2  \ .
\label{LVIV17}
\end{equation}

The above power-law predator-prey  model is described by two coupled differential equations which do not admit any analytical solution, except for the special previously examined case corresponding to $\alpha_i=1/2$.
The model is form invariant describing power-law interaction also in its Hamiltonian version and belongs to the class of models described by Eqs (\ref{LVII1}) and (\ref{LVII2}). As a consequence, compared to other more general and complex power-law models considered in the literature \cite{SaVo,Savageau,Voit,VoitSavageau,Brenig}, it has the advantage of admitting a Hamiltonian formalism with its first integral easily obtained in closed form. Last but not least, this simple model represents a two-parameter interpolation between the standard LV model ($\alpha_i = 1$) and the model ($\alpha_i = 1/2$), proposed in the previous section, which shows the solution in closed form.

\section{$N$-Component models}

The interaction between $ N $ distinct populations within the LV model is described by means of the following $ N $ coupled nonlinear differential equation
\begin{eqnarray}
\frac{d\,x_i}{d\,t}= \lambda_i \, x_i + x_i\, \sum_{j=1}^N \, A_{ij}\, x_j \, ,
\label{LVV1}
\end{eqnarray}
with $i=1, 2, ... , N$, while the bilinear term $A_{ij}x_i x_j$, represents the competition between the populations $x_i$ and $x_j$. An important generalization of the above $N$-component model was proposed in Ref. \cite{Brenig}, involving power-law functions, whose solution can be obtained by employing numerical integration methods.

Hereafter, we briefly consider the extension of the previously introduced power-law model to $N$-component systems. The evolution equations of the system read
\begin{eqnarray}
\frac{d\,x_i}{d\,t}= \lambda_i \, x_i^{\alpha_i} + x_i^{\alpha_i}\, \sum_{j=1}^N \, A_{ij}\, x_j^{\alpha_j} \ , \
\label{LVV2}
\end{eqnarray}
which can be obtained, starting from Eq. (\ref{LVV1}), describing the LV model, by performing the $x_i \rightarrow x_i^{\alpha_i}$ substitution, on its right-hand side.

In the following we are not interested in the Hamiltonian formalism of the power-law competition system, but instead we focus on the transformation $z_i=\,x_i^{1-\alpha_i}$, which  preserves the power-law form of the system. The evolution equations for the transformed system simplifies to
\begin{eqnarray}
(1+\beta_i)\, \frac{d\,z_i}{d\,t}= \lambda_i + \sum_{j=1}^N \, A_{ij}\, z_j^{\beta_j} \ ,  \ \ \
\label{LVV4}
\end{eqnarray}
with $\beta_i=\alpha_i/(1-\alpha_i)$. The above equations have a form suitable for numerical integration.

It is remarkable that, as in the previously considered case of two-component systems, the $N$-component systems, for $\alpha_i=1/2$ which implies $\beta_i=1$, also admit analytical solutions, in terms of known elementary functions. The evolution equations of the real system become
\begin{eqnarray}
\frac{d\,x_i}{d\,t}= \lambda_i \, \sqrt{x_i} + \sqrt{x_i}\, \sum_{j=1}^N \, A_{ij}\, \sqrt{x_j} \ , \ \
\label{LVV6}
\end{eqnarray}
while the system of the equations governing the transformed system becomes linear i.e.
\begin{eqnarray}
 \frac{d\,z_i}{d\,t}= \frac{1}{2}\, \lambda_i + \frac{1}{2}\,\sum_{j=1}^N \, A_{ij}\, z_j \, ,  \ \ \
\label{LVV7}
\end{eqnarray}
so that its analytical solution is easy to obtain. Finally, the transformation $x_i=\,z_i^{2}$  permits  us to obtain the analytical and explicit solution, of the real power-law competition model described by Eq. (\ref{LVV6}).

\section{Concluding Remarks}

Within the general class of Hamiltonian predator-prey models described by Eqs. (\ref{LVII1}) and (\ref{LVII2}) an integrable model has been identified. This  model described by Eqs. (\ref{LVIII7}) and (\ref{LVIII8}), captures the main features of the standard LV model with the important advantage of admitting the exact analytical solution given by Eqs. (\ref{LVIII25}) and (\ref{LVIII26}).

Two extensions of this model have been considered. The first one regards its generalization to the case of two-component systems with arbitrary power-law competition rates, which leads to the evolution equations in the form given by Eqs. (\ref{LVIV9}) and (\ref{LVIV10}). The second extension of the model regards $N$-component systems governed by evolution equations (\ref{LVV2}).

Future perspectives and developments of the above models may be related to both their application to specific fields of science where the LV model is usually employed, as well as to their further extensions. Regarding the possible extensions of the here considered models, one can wish and welcome their future stochastic, fractional and diffusive generalizations. Furthermore, by modifying some of the rate terms or by adding new rate terms to the evolution equations of the models, it may be possible to treat specifical empirically observed phenomenologies (Allee, Holling, etc).

\end{document}